\documentclass[aip,graphicx]{revtex4-1}

\usepackage{graphicx}
\usepackage{amsmath}
\usepackage{amssymb}
\usepackage{dcolumn}
\usepackage[utf8]{inputenc}
\usepackage[T1]{fontenc}
\usepackage[colorlinks,citecolor=red,urlcolor=blue,bookmarks=false,hypertexnames=true]{hyperref}
\usepackage{multirow}
\usepackage{xcolor}
\newcommand{\add}[1]{{\color{black}{#1}}} 
\newcommand{\addd}[1]{{\color{black}{#1}}} 
\newcommand{\prevadded}[1]{{#1}} 
\usepackage{lineno}
\begin{document}


\title{Role of Micellar Entanglements on Kinetics of Shear Banding Flow Formation}
\author{Peter Rassolov}






\author{Hadi Mohammadigoushki}
\email[]{hadi.moham@eng.famu.fsu.edu}
\affiliation{Department of Chemical and Biomedical Engineering, FAMU-FSU College of Engineering, Tallahassee, FL 32310, USA.}


\date{\today}

\begin{abstract}
We investigate the effects of micellar entanglement \prevadded{number} on the kinetics of shear banding flow formation in a Taylor-Couette flow. Three sets of wormlike micellar solutions, each set with a similar fluid elasticity and zero-shear-rate viscosity, but with varying entanglement densities, are studied under start-up of steady shear. Our experiments indicate that in the set with the low fluid elasticity, the transient shear banding flow is characterized by the formation of a transient flow reversal in a range of entanglement densities. Outside of this range, the transient flow reversal is not observed. For the sets of medium and high elasticities, the transient flow reversals exist for relatively small entanglement densities, and disappear for large entanglement densities. Our analysis shows that wall slip and elastic instabilities do not affect the transient flow feature.  We identify a correlation between micellar entanglement \prevadded{number}, the width of the stress plateau, and the extent of the transient flow reversal. As the micellar entanglement \prevadded{number} increases, the width of the stress plateau first increases, then, at a higher micellar entanglement \prevadded{number}, plateau width decreases. Therefore, we hypothesize that the transient flow reversal is connected to the micellar entanglement \prevadded{number} through the width of the stress plateau.
\end{abstract}

\maketitle
\section{Introduction and background}
\label{scn:intro-and-bkg}

Surfactants and salts in aqueous solutions can self-assemble into micelles of different shapes, and among them are wormlike micelles (WLMs), which are long flexible cylindrical assemblies of surfactant molecules. WLMs that form entangled networks exhibit strong viscoelastic properties that make them useful in various practical applications such as in personal care products\cite{Chu2013}, oil-gas fields\cite{Zana2007}, and as model viscoelastic systems for fundamental research\cite{Cates2006-review,Rothstein2020-review}. In particular, the viscoelastic properties of WLMs can be tuned as needed by adjusting the types and concentrations of the surfactants and salts. \par

One common and well-studied flow phenomenon observed in some WLMs is shear banding. Shear banding is associated with the formation of (at least) two co-existing bands of different shear rates that undergo the same shear stress~\cite{Divoux2015-shearbanding-review,Spenley1996-first-nonmonotonic}. Our understanding of the shear banding in WLMs is advanced in Taylor-Couette (TC) flows (flow between two concentric cylinders)\cite{Divoux2015-shearbanding-review,Manneville2008-shearbanding-review,Olmsted2008-shearbanding-review}. A key aspect of the shear banding flows that has been studied is the kinetics of shear banding flow formation upon inception of flow; i.e., the temporal evolution of flows before a quasi-steady shear banded flow is established\cite{Al-kaby2018-rheonmr-shearbanding,Al-kaby2020-rheonmr-shearbanding,Becu2004-wallslip-shearbanding,Brown2011-rheonmr-wallslip,Davies2010-velocimetry-shearbanding,Hu2005-shearbanding-evolution,Lettinga2009-wallslip-shearbanding,Miller2007-transientflow-wallslip,Mohammadigoushki2019-softmatter}. Recent studies in TC flows have shown that the kinetics of shear banding flow formation in WLMs feature wall slip\cite{Becu2004-wallslip-shearbanding,Brown2011-rheonmr-wallslip,Fardin2012-instabilities-wall-slip,Feindel2010-fluctuations-wallslip,Kuczera2015-rheonmr-instabilities-wallslip,Lettinga2009-wallslip-shearbanding,Lopez-Gonzalez2004-shearbanding-fluctuations,Miller2007-transientflow-wallslip}, elastic instabilities\cite{Al-kaby2020-rheonmr-shearbanding,Fardin2008-instabilities,Fardin2012-shearbanding-instabilities,Kuczera2015-rheonmr-instabilities-wallslip}, micellar alignment\cite{Helgeson2009,Gurnon2014} and transient flow reversal\cite{Mohammadigoushki2019-softmatter,Rassolov2020}. In principle, the kinetics of the shear banding flow formation can be influenced by two factors; (i) the TC flow geometry and (ii) material properties of the WLMs. The effects of flow geometry, in particular the surface conditions of the flow cell, have been studied fairly extensively\cite{Lettinga2009-wallslip-shearbanding,Lerouge2008-interface-instabilities}. Lettinga and Manneville showed that modification of the surfaces of the TC cell affects the kinetics of shear banding flow formation through enabling or preventing wall slip. Wall slip can interfere with shear banding but can be diminished by roughening the surfaces in the measurement cell\cite{Lettinga2009-wallslip-shearbanding}. In transient flow, Mohammadigoushki and coworkers \cite{Mohammadigoushki2019-softmatter} quantified wall slip in start-up flow for a CTAB/NaSal solution using rheo-PTV and found that wall slip is most prominent at two times: (1) immediately after the start of applied shear, and (2) as the high shear band initially forms. Additional studies have illustrated that as the high shear band initially forms, the interface between the high and low shear bands becomes unstable due to formation of secondary flows in the high shear band near the inner rotating cylinder \cite{Fardin2016-el-instabilities,Lerouge2006-instabilities-interface,Lerouge2008-interface-instabilities,Mohammadigoushki2016-interface-instabilities}.\par

While the effects of TC flow geometry have been studied, less is known about the effects of material properties of WLMs on the kinetics of shear banding flow formation based on experiments. Our group has recently started investigating the role of material properties on the kinetics of shear banding flow formation in experiments. In particular, Mohammadigoushki and coworkers \cite{Mohammadigoushki2019-softmatter} observed a transient flow feature in flows of shear banding CTAB/NaSal WLMs. When a steady shear flow is first applied, there may be a temporary reversal in flow direction in part of the flow field during the formation of the steady-state shear-banding flow, which occurs during the transient stress decay period \cite{Mohammadigoushki2019-softmatter}. By comparing these observations with the predictions of the Vasquez-Cook-McKinley (VCM) model, the authors suggested that these transient flow reversals are associated with a relatively high fluid elasticity $E$\cite{Mohammadigoushki2019-softmatter} compared to prior literature reporting experiments of transient flow evolution of WLMs~\cite{Miller2007-transientflow-wallslip,Lettinga2009-wallslip-shearbanding,Hu2005-shearbanding-evolution}with the system exhibiting transient reversal characterized by $E \sim 10^8$. The fluid elasticity is defined as $E$ = $\mathrm{Wi} / \mathrm{Re}$, where $\mathrm{Wi} = \lambda\dot{\gamma}$, $\mathrm{Re} = \rho \dot{\gamma}d^2/ \eta_0$. Here, $\lambda$ is the fluid relaxation time, $\dot{\gamma}$ is the shear rate, $\rho$ is the fluid density, $d$ is the gap size, and $\eta_0$ is the zero-shear viscosity. According to the predictions of the VCM model, the transient flow reversal occurs only beyond a critical fluid elasticity number, $E$ and a critical dimensionless applied shear ramp-up rate $a$.  The dimensionless shear ramp-up rate $a$ is also defined as $a = \lambda / t_s$, where $t_s$ is the duration of the initial shear ramp~\cite{Zhou2012-VCM-multi-banding,Zhou2014-VCM-predictions}. These predictions were confirmed in a subsequent study by Rassolov and Mohammadigoushki\cite{Rassolov2020}, which reports a critical threshold in $E$ and $a$ beyond which a transient flow reversal is observed. The critical transition in $E$ reported in study by Rassolov and Mohammadigoushki\cite{Rassolov2020} was around $E \sim 10^6$; at such large values of $E$, there may be traveling elastic waves with a time scale much shorter than the fluid relaxation time. The transient flow reversals occur at times longer than elastic waves but still shorter than the fluid relaxation time. \add{However, the speed of the elastic wave was found to increase with an increase in fluid elasticity. Therefore, critical transitions in transient flow reversal with high elasticity may be related to the interaction of these elastic waves with the TC cell boundaries~\cite{Rassolov2020}.}

Besides the VCM model, predictions of transient flow reversals have been reported using the diffusive Rolie-Poly (DRP) model\cite{Adams2011-DRP}. Adams et al. showed that the formation of transient flow reversals in viscoelastic polymer solutions depends on the polymer entanglement \prevadded{number}, $Z$, and the viscosity ratio, $\beta = \eta_s / \eta_p$, where $\eta_s$ is the solvent viscosity and $\eta_p \approx \eta_0 - \eta_s$ is the solute zero-shear viscosity. In our recent experimental study, we found that increasing the value of the micellar entanglement \prevadded{number} $Z$ at a constant fluid elasticity may tend towards transient flow reversal\cite{Rassolov2020}; however, this observation is based on one comparison. A systematic study is needed to fully understand the effect of micellar entanglement \prevadded{number} $Z$ on the transient evolution of shear banding flows. Therefore, the main objective of this study is to assess the effect of $Z$ on this flow feature using both experiments and simulations. In the following sections, we will present a systematic study of a start-up flow evolution for sets of WLMs where the value of $Z$ is varied while those of the other relevant material parameters (particularly $E$ and $\beta$) are held nearly fixed.

\section{Experiments}
    \label{scn-expts}
\subsection{Materials}

Fluids were prepared in the same manner as in our prior work \cite{Rassolov2020}. Cetyltrimethylammonium bromide (CTAB) and sodium salicylate (NaSal) were mixed in deionized water. Both solutes were obtained from Millipore Sigma and used as received. Following mixing, solutions were kept sealed and away from ambient light for a minimum of two weeks prior to any measurements. For rheo-PTV experiments, glass microspheres (Potters 110P8, diameter $\approx 8 \mu \mathrm{m}$) were mixed at 50 ppm by mass along with the solutes. For visualization of flow instabilities, mica flakes (Jacquard PearlEx 671) were added at 250 ppm in a similar manner.

\subsection{Fluid characterization}

To find the relaxation time and the entanglement \prevadded{number} of the fluids, linear viscoelastic data were measured using Small-Amplitude Oscillatory Shear (SAOS) experiments in an Anton-Paar MCR 302 with an off-the-shelf TC measuring cell with dimensions $R_i$ = 13.328 mm, $R_0$ = 14.449 mm, and $h$ = 40 mm. Here $R_i$ and $R_0$ refer to inner and outer radii respectively, and $h$ is the height of the TC cell. To assess the micellar entanglement \prevadded{number}, the local minimum in the loss modulus ($G^{\prime \prime}$) where the elastic modulus ($G^{\prime}$) approaches a plateau is needed (see details below). For several of the selected fluid preparations, the local minimum in the loss modulus occurs at a fairly high frequency, near 100 radians/s. This is close to the high frequency limit accessible in SAOS experiments beyond which measurements have significant errors due to inertia in the measuring system. For fluid preparations where the minimum in $G''$ is at or near this limit, a different method based on Diffusing Wave Spectroscopy (DWS) was used to extend the measured frequency spectrum to higher frequencies. For DWS experiments, fluids were prepared with 1\% by mass of latex spheres (Life Technologies, $R$ = 300 nm) and measured using a commercially available DWS instrument, LS Instruments RheoLab II, in 5 mm cuvettes with a 300 s multi-tau measurement time and a 60 s echo time. A preparation of 1\% latex spheres in \prevadded{deionized} water was used to obtain the mean free path length $l^{*}$ for DWS measurements.\par

To find the zero-shear viscosity and the high and low shear rate limits of shear banding, steady shear stress measurements were completed at shear rates from 0.001 to 100 s$^{-1}$ in the same measurement cell as for SAOS. The viscosity was measured over time under each applied shear rate until it reached a steady value or quasi-steady oscillation, and either the steady value or the average value over several oscillations respectively was used to assemble the flow curve.

\subsection{Rheo-optical measurements}

Rheo-optical measurements were completed using a custom-built TC cell \add{coupled to the MCR 302} with $R_i$ = 13.35 mm, $R_o$ = 14.53 mm, and $h$ = 50 mm. Details of the cell design and operation are given in our previous work \cite{Mohammadigoushki2019-softmatter,Rassolov2020}. Different from our previous work, the inner cylinder in this study was roughened by sandblasting in order to reduce wall slip (\add{See a preliminary investigation of the effects surface types on transient flow reversal, transient and quasi-steady wall slip in section (I) of the supplementary materials, which will be completed in our future work.)}. Additionally, all rheo-optical measurements were completed under initially applied steady shear rate flow with the shortest possible ramp-up duration (measured to be 0.1 s). Spatially and temporally resolved fluid velocity was measured using rheo-PTV. As in our previous work, the fluids were prepared with glass microspheres, and a laser (with wave-length of 532 nm) and a high speed camera (Phantom Miro 310) were used to image the flow plane. Particle trajectories were obtained from short clips of the video of the flow (much shorter than the time scales of flow evolution) using a Python script based on the TrackPy library \cite{Allan2019-trackpy} and (in the same script) averaged over small intervals along the radial axis to obtain spatially resolved velocity data with quantified uncertainty. In a modified version of the script, the above analysis was repeated for many consecutive intervals throughout longer videos to obtain a spatiotemporal map of either start-up or quasi-steady flow velocity.\par

In addition to rheo-PTV, flow instabilities were visualized using the fluids prepared with mica flakes, which orient to the flow direction and reflect light in an orientation-dependent manner. The fluid in the TC cell was illuminated by a desk lamp, and videos of the fluid flow were captured using a USB video camera (SenTech STC-MBS241U3V) aimed along the \prevadded{radial direction} of the TC cell (i.e. imaging the \prevadded{$x_1 -x_3$} plane). A Python script was used to assemble the video frames into a static image showing spatiotemporal evolution of the flow pattern. Further details of such visualizations are give in our previous work\cite{Mohammadigoushki2017-instabilities}. \prevadded{To characterize the flow instabilities quantitatively and resolve them from other sources of variation in light intensity, power spectra were obtained by fast Fourier transformation (FFT) of the images, with the $x_3$ axis transformed to wavenumber $k$.}

\section{Results and discussions}
\label{scn-results}
\subsection{Fluid Selection and Characterization}

Following characterization of the linear and non-linear viscoelastic properties of the micellar solutions, three sets of fluid preparations, each with closely matched $E$, closely matched $\eta_0$, and broadly varied $Z$, were identified. Fig.~\ref{fig:rheo_results} shows the rheological results for these samples, and the sample compositions and the rheological properties are listed in Table~\ref{tbl:fluid_conditions}. \prevadded{The Cole-Cole plots shown in the inset of Fig.~\ref{fig:rheo_results}(a,c,e) exhibit a semi-circular shape indicating that these micellar solutions are in the fast-breaking regime. Additionally, the ratio of the reptation ($\tau_{rep}$) time to the breakage time ($\tau_{br}$), \addd{$\zeta = \tau_{br}/\tau_{rep} \ll 1 $}, thereby, confirming that these systems are in the fast breaking regime (see Table~\ref{tbl:fluid_conditions}).} The micellar entanglement \prevadded{number}, $Z$, cannot be measured directly in experiments. However, previous theoretical studies have provided approximate equations that can be used to assess $Z$ in linear wormlike micellar solutions. Cates and Granek \cite{Granek1992_orig_Z} developed a scaling relationship that links the micellar entanglement \prevadded{number} to the measured storage and loss moduli in the fast breaking regime as $Z_{CG} \sim {(\prevadded{G_\mathrm{N}^{0}}/G''_{min})}$, where $\prevadded{G_\mathrm{N}^{0}}$ and $G''_{min}$ respectively denote the plateau modulus and the local minimum in the loss modulus at high frequencies for which the storage modulus shows a plateau. Later, Granek~\cite{Granek1994} incorporated the effects of fluctuations in contour-length and suggested an improved version of this scaling relation as $Z_G^{0.82}\approx {(\prevadded{{G_\mathrm{N}^{0}}}/G''_{min})}$. More recently, Larson and co-workers used the simulations of their pointer algorithm and through fitting to experimental data for a series of micellar solutions, suggested the following scaling relationship \cite{Tan2021_pointeralgorithm_newcorrelation}:
\begin{figure}
    \centering
    \includegraphics[width=0.95\textwidth]{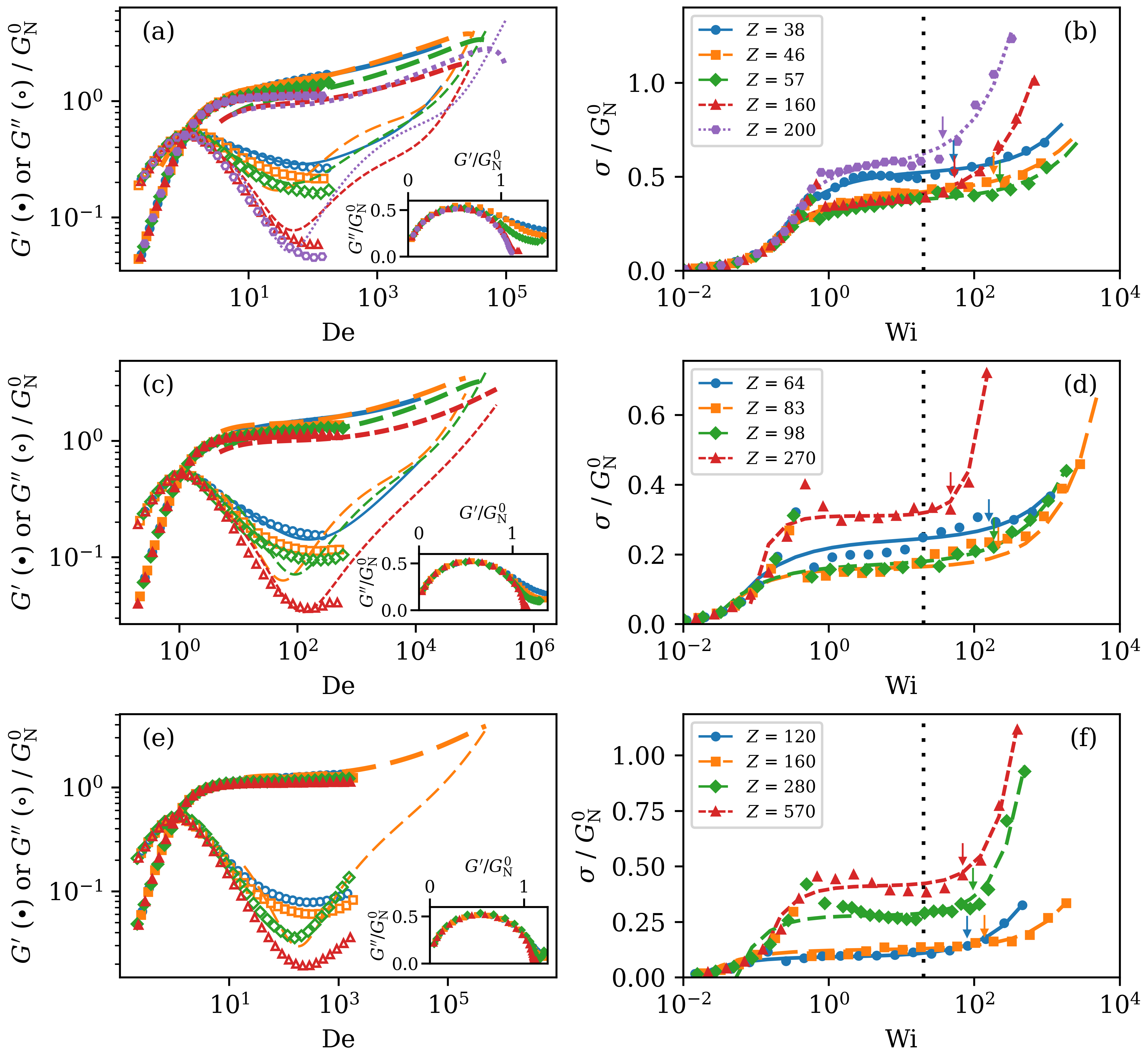}
    \caption{Rheological properties for selected fluid conditions. (a,c,e) Frequency response data shown as SAOS storage modulus (filled symbols) and loss modulus (empty symbols), and adjusted DWS storage modulus (broad curves) and loss modulus (narrow curves) low $E$ (a), medium $E$ (c) and high $E$ (e) sets in Table \ref{tbl:fluid_conditions}. \addd{The Deborah number is defined as $\mathrm{De} = \lambda\omega$.} (a,c,e insets) Cole-Cole representations of the frequency response. (b,d,f) Flow curves for the same conditions as shown in (a,c,e) respectively.\prevadded{ The dotted line indicates Wi~=~20, the applied Wi reported for most of this study, \prevadded{and arrows mark the high shear rate limit of each shear stress plateau}. \add{The dashed curves indicate the best fits of experimental flow curves to a hyperbolic sine function.}}}
    \label{fig:rheo_results}
\end{figure}

\begin{equation}
0.317~Z^{0.82} \approx {(G'_{min}/G''_{min})}
\label{eqn:pointeralgorithm_newcorrelation}
\end{equation}
where $G^{\prime}_{min}$ is $G^{\prime}$ at the same angular frequency as $G^{\prime\prime}_{min}$.

\begin{table}[h!]
    \centering
    \begin{tabular}{ p{1.5cm} p{2.25cm} p{1cm}p{1cm} p{1cm} p{1cm} p{0.75cm} p{0.75cm} p{1cm} p{1.5cm}p{1cm}p{1cm}p{1cm}}
        \hline
        Fluid & CTAB/NaSal & $T$ & $Z_{CG}$ & $Z$ & $\lambda$ & $\tau_{br}$ & $\tau_{rep}$ & $\eta_0$ & $l_p$ & $\langle L \rangle$** & $E$ & $G_\mathrm{N}^0$ \\
              & (mM/mM) & ($^\circ$C) & & & (s) & (ms)* & (s)* & (Pa$\cdot$s) & (nm) & ($\mu$m) & $\times10^{5}$ & (Pa) \\
        \hline\hline
         \multirow{5}{2cm}{Low $E$}
         & 250/45   & 25 & 6.2 &  38 & 1.64 & 13.3 &  201 &  123 &   -  & 1.7 & 1.45 & 74.6 \\
         & 250/50   & 28 & 7.4 &  46 & 1.47 & 10.0 &  217 &  130 & 31.3 & 1.9 & 1.38 & 86.3 \\
         & 200/50   & 30 & 8.7 &  57 & 1.78 & 13.3 &  237 &  122 & 30.6 & 2.6 & 1.56 & 69.4 \\
         & 200/70   & 38 & 20  & 160 & 1.20 & 10.0 &  144 &  136 &   -  & 5.9 & 1.17 & 109  \\
         & 150/67.5 & 40 & 25  & 200 & 1.85 & 17.8 &  193 &  133 & 29.9 & 9.6 & 1.77 & 70.9 \\
        \hline
        \multirow{4}{2cm}{Medium\\$E$}
         & 250/45   & 19 & 9.6 & 64  & 6.25 & 23.7 & 1650 &  533 &   -  & 2.6 & 23.9 & 85.3 \\
         & 250/50   & 22 & 12  & 83  & 5.09 & 23.7 & 1090 &  576 &   -  & 3.1 & 21.1 & 103  \\
         & 200/50   & 25 & 14  & 98  & 5.89 & 23.7 & 1470 &  472 & 29.9 & 4.3 & 20.0 & 77.1 \\
         & 200/70   & 32 & 32  & 270 & 4.72 & 31.6 &  704 &  513 &   -  & 10  & 19.2 & 107  \\
        \hline
        \multirow{4}{2cm}{High $E$}
         & 250/50   & 17 & 16  & 120 & 14.5 & 42.2 & 5010 & 1670 &   -  & 4.4 & 175  & 107  \\
         & 200/50   & 20 & 20  & 160 & 18.4 & 56.2 & 6060 & 1540 &   -  & 6.8 & 204  & 80.6 \\
         & 200/70   & 27 & 32  & 280 & 15.7 & 100  & 2450 & 1600 &   -  & 11  & 180  & 102  \\
         & 150/67.5 & 29 & 57  & 570 & 22.0 & 100  & 4820 & 1590 &   -  & 27  & 251  & 69.2 \\
        \hline
    \end{tabular}
    \caption{List of shear banding wormlike micellar solutions used in this study and the rheological characteristics. $Z_{CG}$ is the estimate of the entanglement \prevadded{number} according to the Cates and Granek scaling law, and $Z$ is the estimate according to the Larson scaling relation.
    * As in our previous study~\cite{Rassolov2020}, the breakage time is obtained from the expression: $\tau_{br} \approx 1/\omega_{min}$ and the reptation time from: $\lambda \approx \sqrt{\tau_{br}\tau_{rep}}$ where $\omega_{min}$ is the angular frequency at $G''_{min}$. \addd{Where values of $\tau_R$ can be reliably obtained from DWS data, they can also be used to estimate the micellar persistence length, $l_p$, by the expression $\tau_R^{-1} \approx \frac{k_B T}{8 \eta_s l_p^3}$\cite{Oelschlaeger2009-dws-details} where $\eta_s$, the viscosity of water, is estimated from an empirical correlation\cite{Huber2009}. The persistence length can be used in another correlation to obtain the entanglement length $l_e$: $G_\mathrm{N}^0 = \frac{k_B T}{{l_e}^{9/5}{l_p}^{6/5}}$\cite{Oelschlaeger2009-dws-details,OMIDVAR201848}, and finally, it is possible to estimate the average length of the micelles using the definition $Z = \langle L \rangle / l_e$. ** For fluid preparations where $l_p$ cannot be obtained reliably from DWS data and is not reported, 30 nm is used as an approximation. The persistence length of CTAB/NaSal/water WLMs is between 29 and 45 nm for a broad range of salt/surfactant ratios and temperatures\cite{GalvanMiyoshi2008}.}}
    \label{tbl:fluid_conditions}
\end{table}

Eq.~(\ref{eqn:pointeralgorithm_newcorrelation}) differs from the scaling relation of Granek in that the prefactor of unity and the plateau modulus are replaced with a prefactor of 0.317 and the storage modulus at the frequency for which loss modulus shows a local minimum. In our study, we primarily use Larson's relation to estimate the micellar entanglement \prevadded{number}; however, we also report the estimate according to Granek's relation. In order to do so, the rheological properties $G'_{min}$ and $G''_{min}$ are needed for the former, and $\prevadded{{G_\mathrm{N}^{0}}}$ and $G''_{min}$ are needed for the latter. For our selected fluid preparations, $\prevadded{{G_\mathrm{N}^{0}}}$ can be \prevadded{estimated as $G'_{min}$}. However, for some WLMs, $G''_{min}$ and $G'_{min}$ are not captured within the range of frequencies accessible by this method ($\omega \le 100$ rad/s). To obtain higher frequency linear viscoelastic response of such fluids, we used DWS. DWS results were fitted to mechanical rheometry results by multiplying the storage and loss moduli by an optimized prefactor. This fitting procedure and its theoretical justification are described in detail by Larson and co-workers \cite{Zou2019_pointeralgorithm_dws} and are plotted as the dashed lines in Fig.~\ref{fig:rheo_results}(a,c,e). The DWS method is only reliable for frequencies beyond 10 rad/s and is known to deviate from the mechanical rheology data for which $G'$ is of much larger magnitude than $G''$ \cite{Oelschlaeger2009-dws-details}. Therefore, accurate and reliable DWS results could not be obtained for WLMs with high entanglement \prevadded{number}, and hence, DWS results are not plotted for such fluid preparations. However, $G'_{min}$ and $G''_{min}$ for these particular fluid preparations can be estimated reliably from mechanical rheometry data alone. \prevadded{Additionally, there is a shoulder evident around $\lambda\omega \approx 10^3 - 10^4$ for some of the DWS spectra, and it is not currently clear what may have caused it. One possibility is that with the high viscosity (requiring long settling times) and relatively high ionic strength of the selected WLM solutions, some particle aggregation and adsorption to the cuvette walls may occur. Nonetheless, $G''_{min}$ occurs at frequencies well below this shoulder. }

\addd{In addition to Larson's correlation, Peterson and Cates\cite{Peterson2020full} report a scaling law relating $Z$ to relevant time scales as $\lambda / \tau_R \sim Z \zeta^{1/2}$, where $\tau_R$ is the Rouse time. The Rouse time can be estimated as $\tau_R \approx 1/\omega_{c,h}$, where $\omega_{c,h}$ is the higher frequency crossover of $G'$ and $G''$. Some DWS spectra in Fig.~\ref{fig:rheo_results} feature such a crossover, and we have found that the relation between $Z$ and the relevant time scales approximately follows this scaling law (see Fig.~S5 in the supplementary materials). However, for most of the fluid preparations reported, the aforementioned shoulder and difficulties in measuring reliable DWS spectra for high $E$ are likely to introduce large errors into any attempt to estimate $Z$ in this way, and a rigorous test of the scaling law by Peterson and Cates would require further experiments yielding accurate high-frequency data. Therefore, we use Larson's correlation to estimate $Z$ for this study.}

\prevadded{The flow curves of the wormlike micellar solutions are shown in Fig.~\ref{fig:rheo_results}(b,d,f). These systems exhibit a stress plateau within a range of shear rates, as is expected for shear banding systems. \add{Note that these flow curves are obtained using the off-the-shelf TC measuring cell with smooth cylinders. We have repeated these measurements using roughened inner and outer cylinders and did not observe significant changes in the flow curves (see Fig.~S3 of the supplementary materials for a comparison of smooth cylinders versus roughened cylinders.)} For the low elasticity fluid (Fig.~\ref{fig:rheo_results}(b)), as the micellar entanglement number increases, the width of the stress plateau first increases. However, for $Z>57$, \add{the} stress plateau width shortens as the entanglement number increases. For fluids with a medium elasticity, this critical transition occurs beyond $Z>98$ (see Fig.~\ref{fig:rheo_results}(d)). Finally, for fluids with high $E$, the width of the stress plateau decreases as the micellar entanglement number increases. \add{We have estimated the onset and end of the stress plateau by fitting a hyperbolic sine function to the measured flow curve data (see more details in section (II) of the supplementary materials.).} Previous studies of the Rolie-Poly model have shown that the width of the stress plateau grows monotonically with increasing $Z$ at a fixed solvent viscosity\cite{Adams2011-DRP}. \prevadded{We will come back to this point in subsection~\ref{ssn:plateau-width}.}}

\subsection{\prevadded{Effect} of the imposed Weissenberg number}

\begin{figure}[htp]
    \centering
    \includegraphics{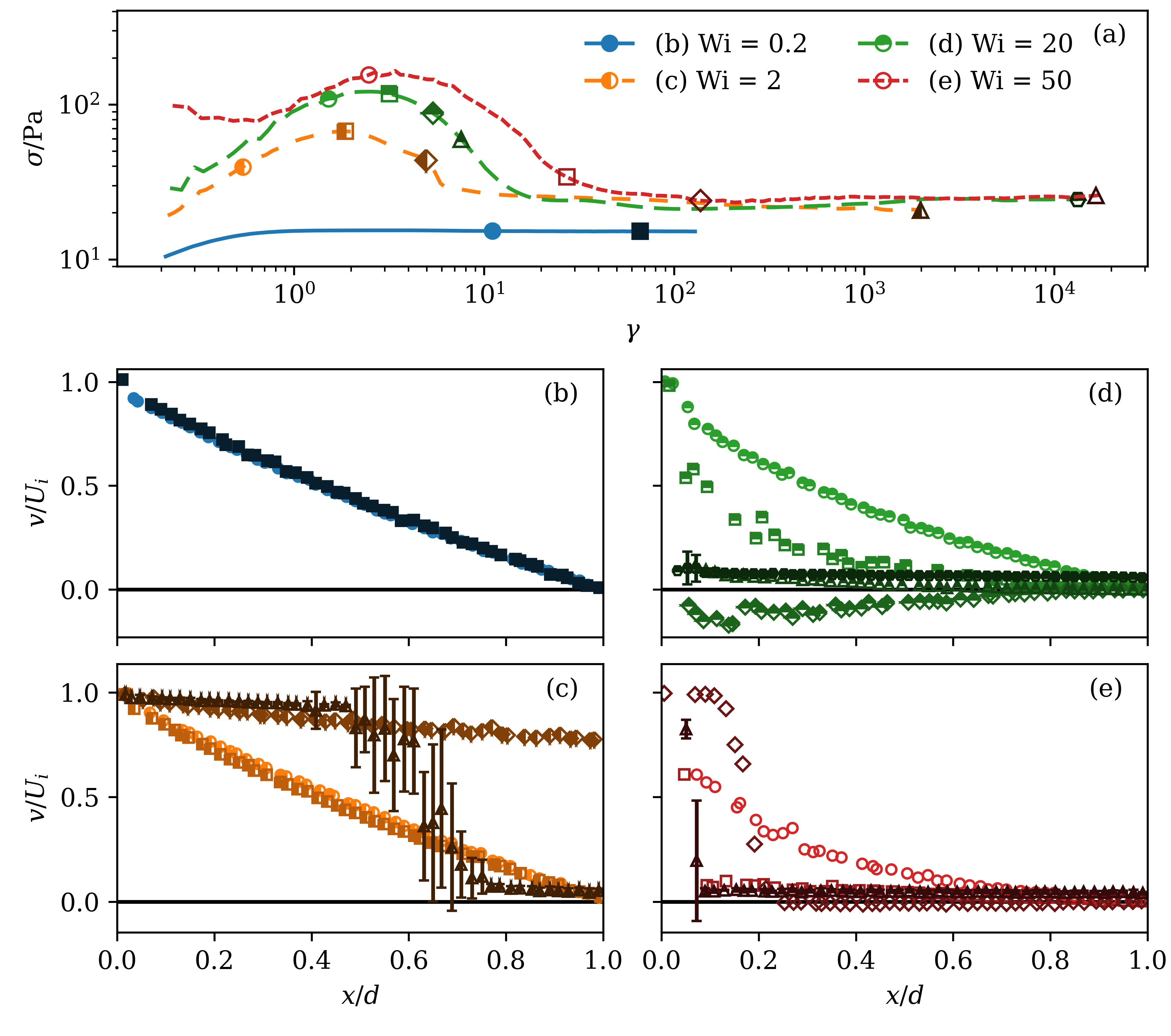}
    \caption{Transient flow evolution for $Z = 57$, $E = 1.56 \times 10^5$ at different Wi. (a) Shear stress evolution with shear strain. (b-e) Temporal evolution of normalized velocity profiles measured in the gap of the TC cell at selected shear strains. Symbols used in (b-e) correspond to those plotted on the shear stress curves in (a), and $U_i$ is the applied velocity at the inner cylinder. \add{Error bars show standard deviations of the quasi-steady velocity.}}
    \label{fig-different-wi}
\end{figure}

As mentioned above, the main objective of this study is to assess the impact of micellar entanglement \prevadded{number} on transient flow reversal, and wall slip may obscure the true impact of the micellar entanglement on this flow feature. The transient and quasi-steady wall slip both depend on the imposed Weissenberg number (Wi), as has been reported in the literature~\cite{Fardin2012-instabilities-wall-slip,Lettinga2009-wallslip-shearbanding}. Therefore, before addressing the impact of micellar entanglement \prevadded{number}, we assessed the impact of the applied Wi on the transient flow reversal in prepared fluids and identified the applied Wi that produced the maximum transient flow reversal (or equivalently the minimum wall slip). \prevadded{Adams et al.\cite{Adams2011-DRP} have shown that the degree of instability of an unstable mode in the DRP model varies with Wi. Based on their calculations, the degree of instability ($\Omega_{max}$) increases as Wi increases, and shows a maximum at the middle towards the end of stress plateau before it decreases again at higher Wi. Although $\Omega_{max}$ does not have an analog in experiments, we expect that the maximum transient flow reversal to occur for a Wi with the strongest instability in the stress plateau region.} Fig.~\ref{fig-different-wi} shows the temporal evolution of the shear stress and velocity profiles for a selected fluid preparation at different applied Wi. Below the onset of shear banding, the transient flow reversal is not observed. For values of Wi that correspond to the onset of shear banding ($\mathrm{Wi} \approx$ 2-5) the transient flow reversal is not observed due to significant wall slip at the outer cylinder. At high Wi (50 and beyond) the transient flow reversal is very weak. Interestingly, at the intermediate values ($\mathrm{Wi} = 20$), wall slip at the outer cylinder is minimal, and the transient flow reversal is at its maximum. This result is also confirmed for other fluid preparations (see Fig. S6 in the supplementary materials), \prevadded{and is qualitatively consistent with the above notion proposed by Adams et al.\cite{Adams2011-DRP} for the DRP model.} Therefore, for the remainder of this work, we focus on the impact of micellar entanglement \prevadded{number} at a fixed Weissenberg number $\mathrm{Wi} = 20$ for which the transient flow reversal is at its maximum.\par

\subsection{Effects of \prevadded{the} micellar entanglement \prevadded{number}}

\begin{figure}[htp]
    \centering
    \includegraphics{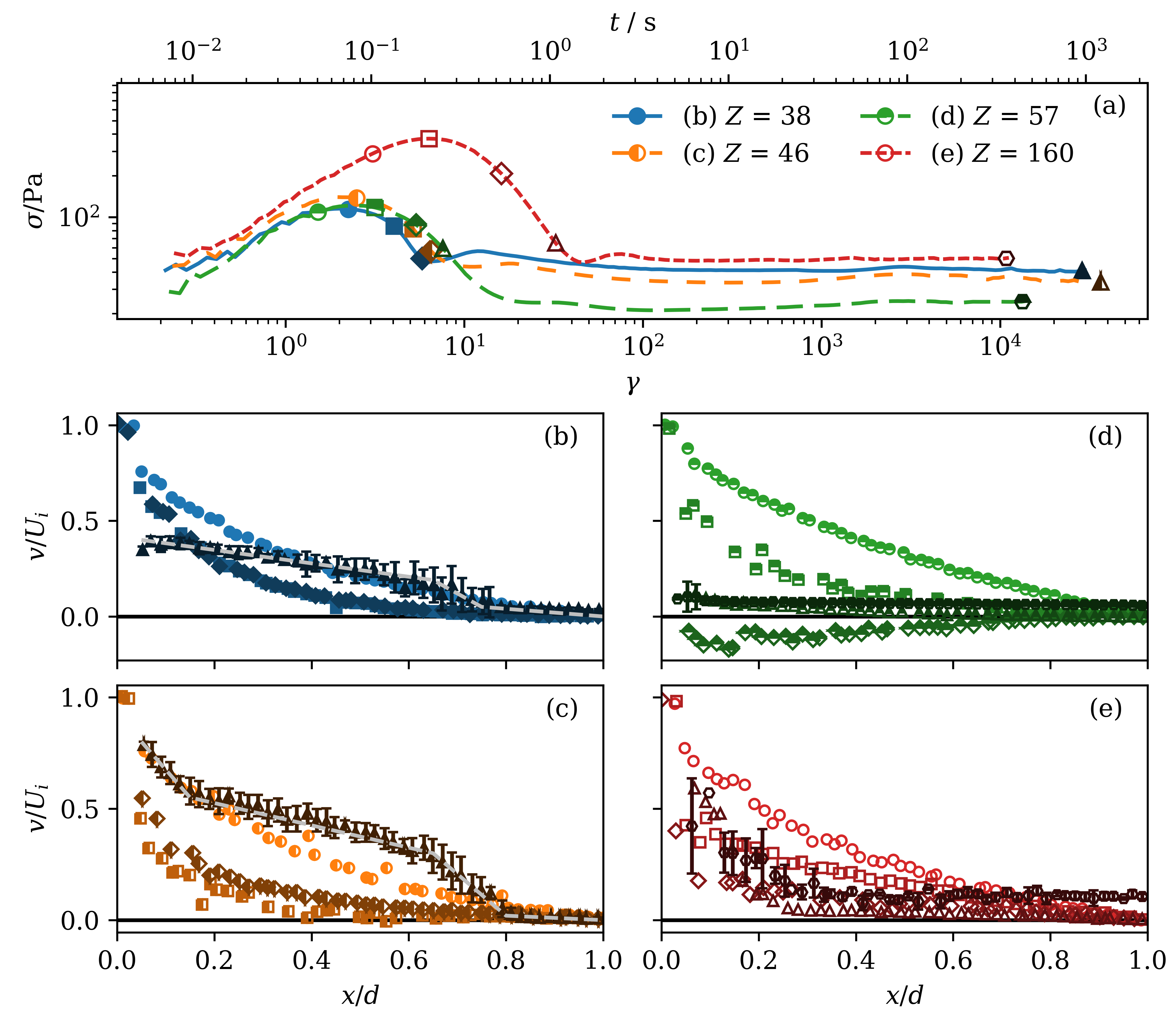}
    \caption{Transient flow evolution for $E = (1.36 \pm 0.20) \times 10^5$ and Wi = 20 at varied $Z$. (a) Shear stress evolution with shear strain. \add{The accumulated shear strain $\gamma$ is reported by the MCR 302, and the} time $t$ is computed from $\gamma$ using the geometric mean of the shear rates in each of (b-e). (b-e) Temporal evolution of normalized velocity profiles measured in the gap of the TC cell at selected shear strains. Symbols used in (b-e) correspond to those plotted on the shear stress curves in (a), and $U_i$ is the applied velocity at the inner cylinder. Dashed lines in (b,c) are visual aids showing quasi-steady multiple band profiles. \add{Error bars show standard deviations of the quasi-steady velocity.}}
    \label{fig:low-e-summary}
\end{figure}

Fig. (\ref{fig:low-e-summary}) shows transient evolution of the flow for the set of micellar solutions with low $E$ at $\mathrm{Wi} = 20$ for various entanglement densities. For the lowest $Z$ in this set, $Z = 38$, flow develops inhomogeneity (i.e., deviates from a linear velocity profile) during the stress overshoot, and this inhomogeneity then evolves to a shear banded profile. However, the elastic recoil is not strong enough to generate transient flow reversal. At longer times, a multiple-band quasi-steady velocity profile forms, which is characterized by \add{three visible shear bands forming a low-high-low band order from $x/d \approx 0.05$ toward the outer cylinder. No clear particle images are observed for $x/d < 0.05$; this region may feature either a high shear band or a continuation of the nearer low shear band with strong wall slip.}
As $Z$ increases slightly to 46, the flow inhomogeneity becomes more significant \add{and a high shear band near the inner cylinder becomes clearly quantifiable in the quasi-steady flow}; however, the transient and quasi-steady flows are otherwise unchanged. At a still higher micellar entanglement \prevadded{number} ($Z = 57$), the flow features a transient reversal in direction. A similar flow feature has been reported in prior literature for $E \sim 10^{7}$~\cite{Mohammadigoushki2019-softmatter,Rassolov2020}; however, it had not been observed for such a low elasticity as $E = 1.56 \times 10^5$. Interestingly, as $Z$ is increased to 160, the transient flow reversal disappears.\begin{figure}[htp]
    \centering
    \includegraphics{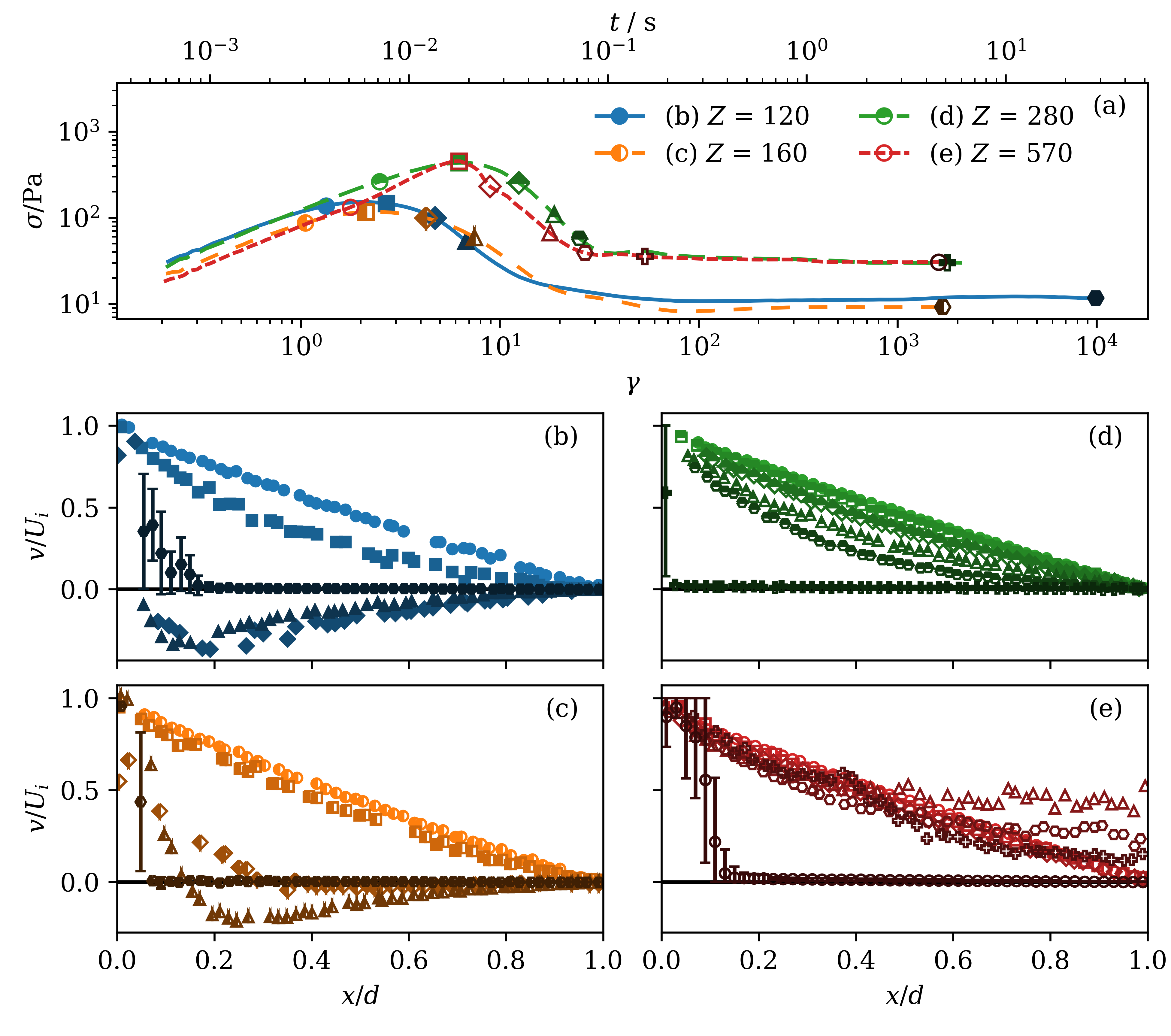}
    \caption{Transient flow evolution for $E = (2.13 \pm 0.38) \times 10^7$ and Wi = 20 at varied $Z$. (a) Shear stress evolution with shear strain. \add{The accumulated shear strain $\gamma$ is reported by the MCR 302, and the} time $t$ is computed from $\gamma$ using the geometric mean of the shear rates in each of (b-e). (b-e) Temporal evolution of normalized velocity profiles measured in the gap of the TC cell at selected shear strains. Symbols used in (b-e) correspond to those plotted on the shear stress curves in (a), and $U_i$ is the applied velocity at the inner cylinder. \add{Error bars show standard deviations of the quasi-steady velocity.}}
    \label{fig-hi-e-summary}
\end{figure} Similar experiments were performed on the solutions with higher fluid elasticity. Fig.~\ref{fig-hi-e-summary} shows the transient evolution of flow for the high $E$ set. At the lowest micellar entanglement \prevadded{number} from this fluid set ($Z = 120$), the flow features transient reversal followed by a quasi-steady two-banded flow. At higher entanglement densities ($Z = 160$), a similar behavior is reported.  Beyond a critical micellar entanglement \prevadded{number} ($Z$ increased to 280), the transient flow reversal is no longer observed, and the quasi-steady flow features a sharp high shear band near the inner cylinder. A similar transition is reported for fluid sets with medium elasticity (see Fig. S7 in the supplementary materials).\par

Fig.~(\ref{fig:summary}) shows a summary of experimental results in terms of the transient and final quasi-steady flow response \prevadded{at Wi =20}. For the fluids in the set with low elasticity, we observe two transitions in the kinetics of shear banding flow formation depending on the micellar entanglement \prevadded{number}. The first transition is characterized by the appearance of transient flow reversal, while beyond a second critical threshold of micellar entanglement \prevadded{number}, flow reversal is not observed. For fluids with higher elasticity, we only report the second transition. Moreover, Fig.~\ref{fig:summary}(b) shows that the quasi-steady flow profile is characterized by multiple banded structures at the low fluid elasticity set and small micellar entanglement \prevadded{number}. As the micellar entanglement increases beyond a critical threshold, the quasi-steady flow profile is characterized by a two-banded profile. At higher fluid elasticities, the quasi-steady flow forms a two-banded profile regardless of the micellar entanglement \prevadded{number}. \prevadded{Note that the results of Fig.~(\ref{fig:summary}) are only valid for Wi = 20 and the transitions may shift at different imposed Wi.}

The first transition observed in Fig.~\ref{fig:summary}(a) is consistent with existing predictions of the DRP model\cite{Adams2011-DRP}. According to the simulations of the DRP model, viscoelastic polymer solutions exhibit the transient flow reversal beyond a critical polymer entanglement \prevadded{number} (see more detailed discussion below). However, to the best of our knowledge, the second transition observed in Fig.~\ref{fig:summary}(a) is not reported in any model predictions.  \par
\begin{figure}
    \centering
    \includegraphics{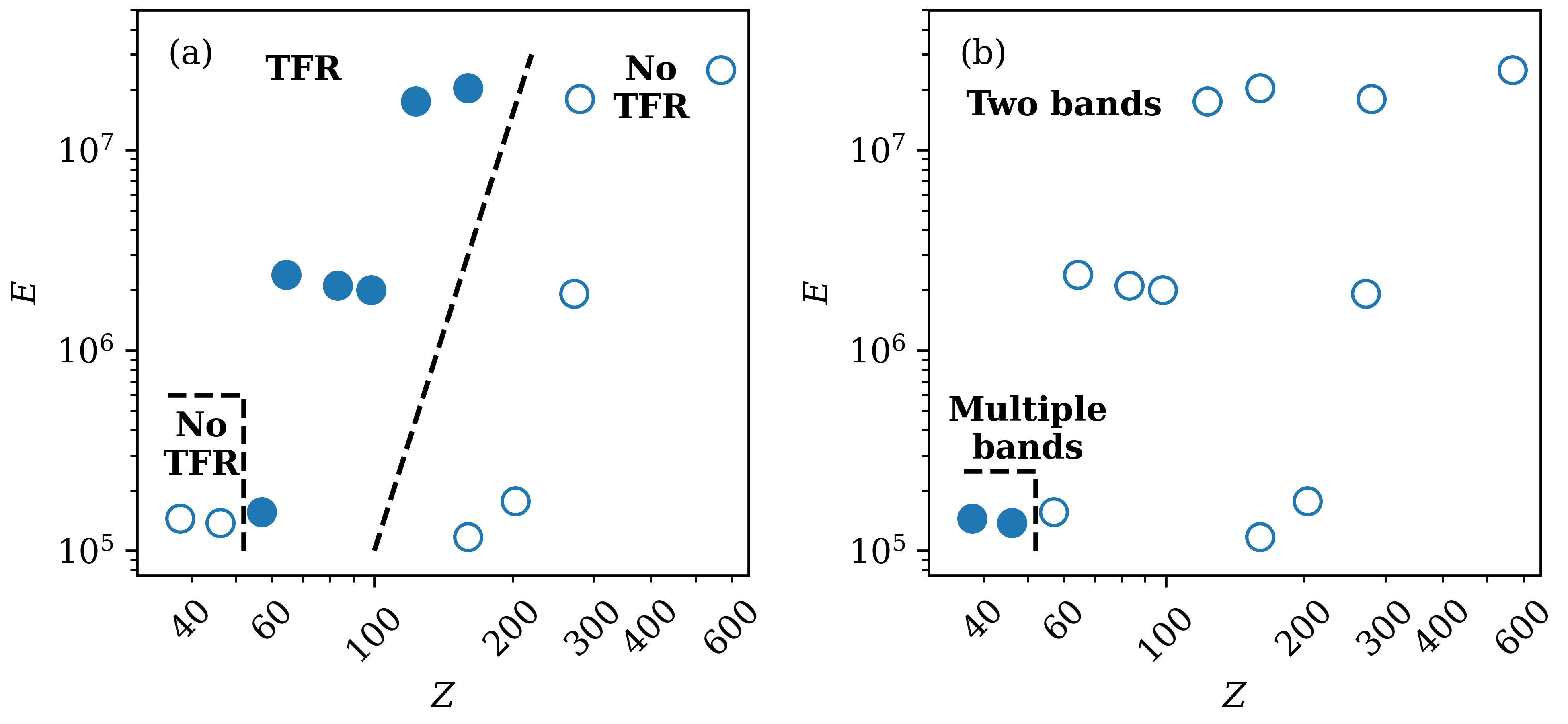}
    \caption{\prevadded{Plots summarizing transient flow reversals and multiple banding in $E$, $Z$ parameter space at Wi = 20. }(a) Filled symbols indicate flow conditions that show transient flow reversals, while empty symbols indicate those that do not. (b) Filled symbols indicate flow conditions that lead to multiple-banded quasi-steady-state flow profiles, while empty symbols indicate those that are characterized by two-banded profiles. }
    \label{fig:summary}
\end{figure}

\subsection{Wall slip and flow instabilities}

As noted in the above, flow of shear banding WLMs features wall slip as well as elastic instabilities. In this section, we will assess any possible connection between the second transition observed in Fig. \ref{fig:summary}(a) and the observed wall slip and/or elastic instabilities.\par
\begin{figure}[htp]
    \centering
    \includegraphics[width=0.9\textwidth]{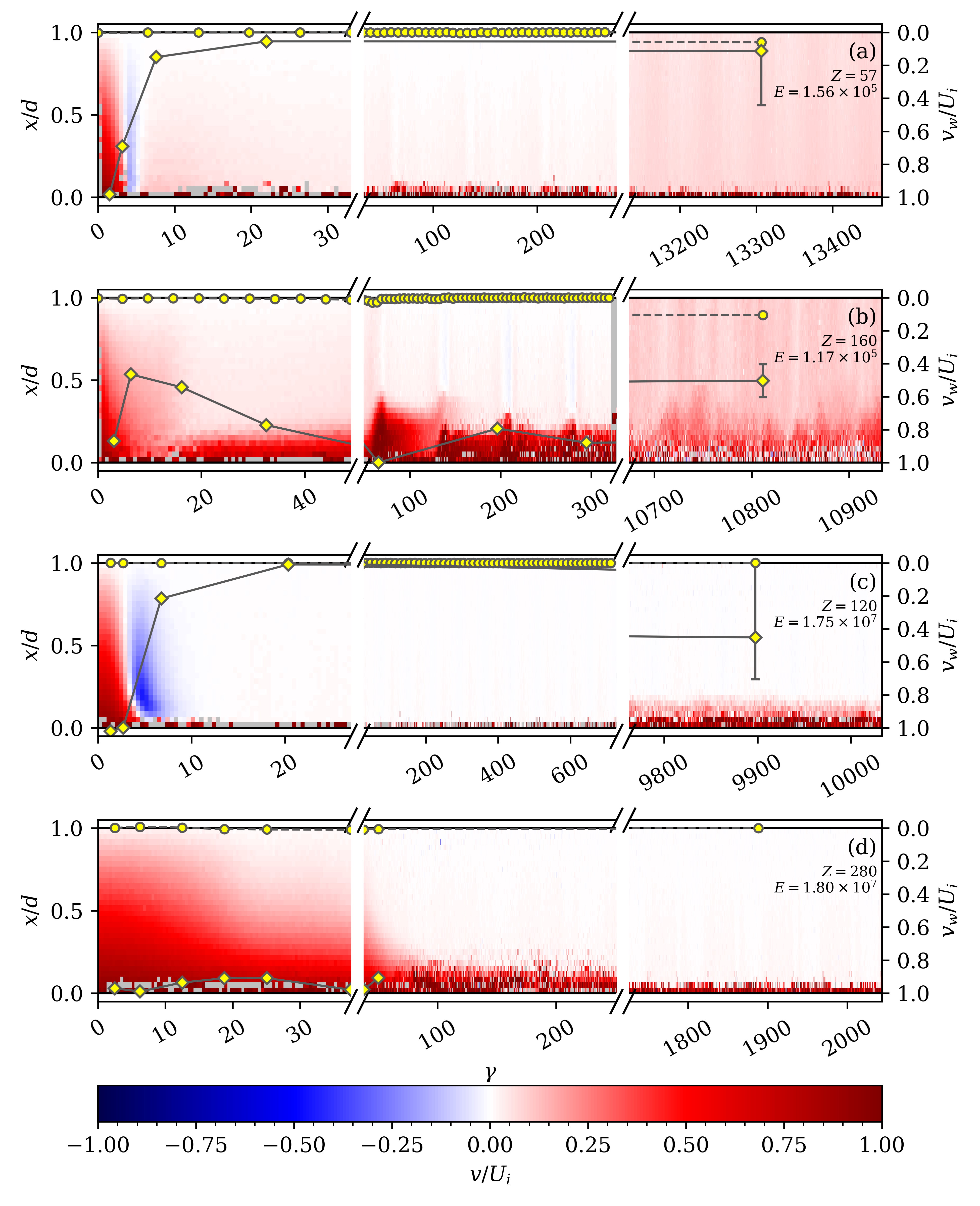}
    \caption{\add{Transient wall slip and velocity maps for selected fluid preparations:} (a) $Z = 57$, $E = 1.56 \times 10^5$. (b) $Z = 160$, $E = 1.17 \times 10^5$. (c) $Z = 120$, $E = 1.57 \times 10^7$. (d) $Z = 280$, $E = 1.80 \times 10^7$. \add{Velocity at the inner cylinder is shown by gray and yellow diamonds and solid lines, while velocity at the outer cylinder is shown by circles and dashed lines. Lines are a visual aid.} \add{For the fluid preparation in (d), optical distortions near the inner cylinder emerge around $\gamma = 50$, and quantification of wall slip at the inner cylinder is not possible.}}
    \label{fig:pvt-wallslip}
\end{figure}
Fig.~\ref{fig:pvt-wallslip} shows the transient wall slip at the inner and outer cylinders for start-up shear flow at $\mathrm{Wi} = 20$ for two of the fluid preparations from the low elasticity set and two from the high elasticity set. Spatiotemporal maps of the transient, local fluid velocity are presented in Fig.~\ref{fig:pvt-wallslip} for each of these four cases to provide a more complete visualization of the transient and quasi-steady flows. \prevadded{To assess the impact of transient wall slip on the second transition from TFR to no TFR reported in Fig.~\ref{fig:summary}(a), we compare the transient wall slip behavior of two pairs of fluid preparations, with each pair having the same fluid elasticity but different entanglement numbers, i.e., $Z = 57$ with $Z = 160$ and $Z = 120$ with $Z = 280$. We start with the pair from the set with the low elasticity.} For these fluids, the flow evolution features transient flow reversal at the lower entanglement \prevadded{number} $Z = 57$ (shown in Fig.~\ref{fig:pvt-wallslip}(a)), while at the higher entanglement \prevadded{number} $Z = 160$, no transient flow reversal is observed (shown in Fig.~\ref{fig:pvt-wallslip}(b)). \add{Transient wall slip emerges at the inner cylinder for both systems and initially increases with increasing shear strain. Following the occurrence or not of transient flow reversal around $\gamma \approx 5$, a high shear band emerges for $Z = 160$, and the wall slip reduces. For $Z = 57$, the shear banding seems to disappear completely in favor of wall slip as shear strain continues to increase (see Fig. ~S8 of the supplementary materials for additional replicates as well as in medium elasticity fluids of Fig.~S7, where a similar behavior is reported).} Finally, in quasi-steady flow \add{for both fluid preparations}, the wall slip at both the inner and the outer cylinders reemerges.
At higher $E$, for $Z = 120$, transient wall slip \add{at the inner cylinder} is observed, while for $Z = 280$, the transient wall slip is much weaker. \add{Under quasi-steady flow, there is transient wall slip at the inner cylinder for $Z = 120$, while for $Z = 280$, optical distortions near the inner cylinder prevent the quantification of wall slip. Therefore, there is likely a connection between the transient evolution of wall slip and the occurrence or not of transient flow reversals. However, to make a decisive determination of how wall slip interacts with the bulk flow and how this may affect the occurrence or not of flow reversals, transient velocimetry with greater spatial resolution than is possible with our apparatus would be needed.}

\begin{figure}
    \centering
    \includegraphics[width=0.9\textwidth]{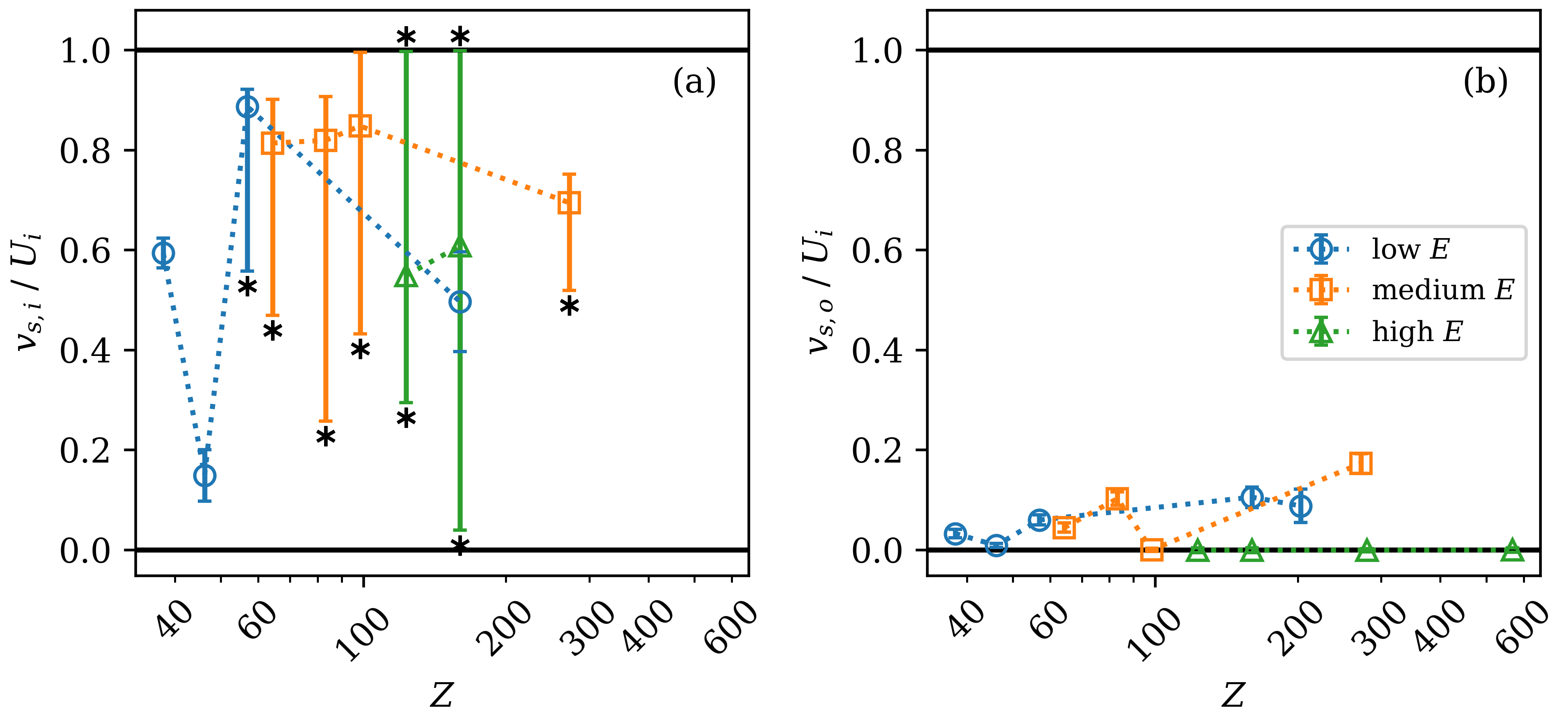}
    \caption{Quasi-steady wall slip for each selected fluid preparation at the inner (a) and outer (b) cylinders, grouped by fluid elasticity $E$. Error bars marked with $*$ indicate the maximum observed extent of slip velocity deviation from the mean, while those without error bars indicate the standard deviation associated with the slip velocity, not exceeding relative slip velocities of 0 or 1.}
    \label{fig:steadywallslip}
\end{figure}
We also monitored the quasi-steady wall slip as a function of entanglement \prevadded{number} for all selected fluid preparations. Fig.~\ref{fig:steadywallslip} shows the quasi-steady wall slip at the inner and outer cylinders as a function of micellar entanglement \prevadded{number}. At the inner cylinder, the quasi-steady wall slip changes in a non-monotonic fashion as the entanglement \prevadded{number} and fluid elasticity are increased. The wall slip measurements have high uncertainty due to temporal fluctuations in the quasi-steady velocity profiles. However, at the outer cylinder, entanglement \prevadded{number} and elasticity affect the observed quasi-steady wall slip. For fluids with the low and medium elasticity sets, the quasi-steady wall slip at the outer cylinder is negligible at low entanglement \prevadded{number} and increases as the entanglement \prevadded{number} increases. However, for the fluids with high elasticity, the quasi-steady wall slip at the outer cylinder is negligible for the entire range of micellar entanglement densities. \par
\begin{figure}[htbp]
    \centering
    \includegraphics[width=0.95\textwidth]{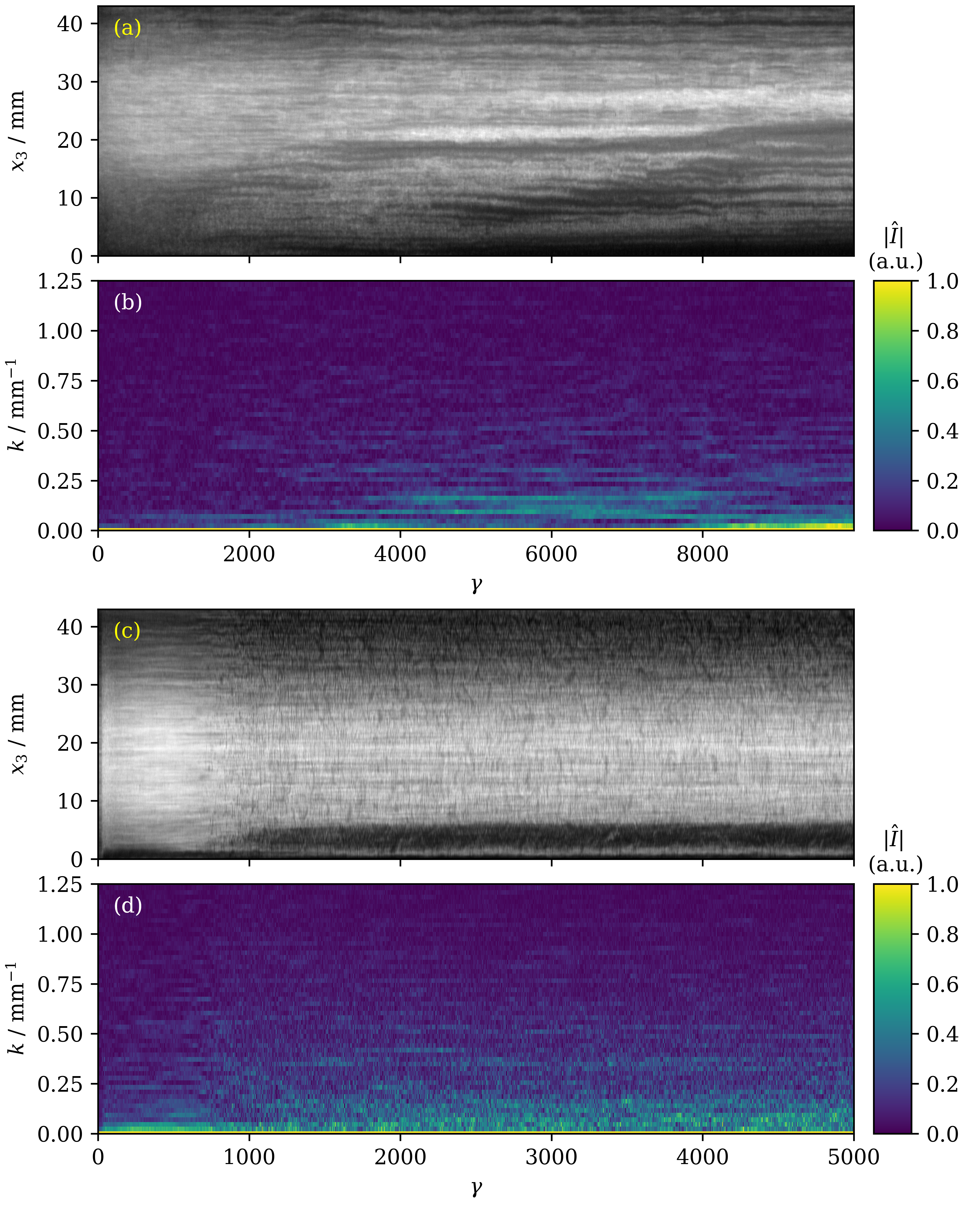}
    \caption{\prevadded{Spatio-temporal evolution of the flow of WLMs subject to start-up of steady shear rate visualized using mica flakes (a,c) and evolution of FFT spectra (b,d) for selected fluid preparations: (a,b) $Z = 57$, $E = 1.56 \times 10^5$. (c,d) $Z = 160$, $E = 1.17 \times 10^5$. Here, $x_3$ denotes the position along the TC cell axis, and $k$ is the wavenumber.}\prevadded{ The onset of flow instabilities is marked with a dotted line.}}
    \label{fig:instabilities}
\end{figure}

Our earlier work~\cite{Rassolov2020} suggested that transient flow reversals in these solutions may be related to elastic instabilities that are known to arise in the flow of some shear banding WLMs \cite{Mohammadigoushki2017-instabilities}. Fig.~\ref{fig:instabilities} shows the evolution of the flow stability in \prevadded{two of the fluids} shown in Fig.~\ref{fig:steadywallslip} subject to the same start-up steady shear flow . These fluids are prepared with mica flakes as described previously to reveal any secondary flows or other instabilities that may occur, which appear as variations in the imaged light intensity in the vorticity direction ($x_3$) \prevadded{in Fig.~\ref{fig:instabilities}(a,c)}. For these fluids, \add{the image of the flow initially features fine striations (wavelength $\lesssim$ 1 mm) in the light intensity that are caused by the individual mica flakes, and darkening of the image towards either end that is caused by variation in the illumination and imaging angle with changing $x_3$ position. Neither of these features indicates flow instabilities; Fig.~S9(a,b) show these features for the same fluid as Fig.~\ref{fig:instabilities}(a) at Wi = 0.2, where no instabilities are expected. Coarser striations, which do indicate instabilities,} arise around $\gamma \approx 3000$ to $4000$ in Fig.~\ref{fig:instabilities}(a) and for $\gamma \approx 1000$ in Fig.~\ref{fig:instabilities}(c). These values of $\gamma$ associated with the onset of instabilities are more clearly evident in the FFT spectra shown in Fig.~\ref{fig:instabilities}(b,d). \add{In the frequency domain, the long-range variation in light intensity appears as low-wavenumber modes ($k \lesssim 0.1$ mm$^{-1}$) and the striations due to the individual mica flakes appear as weak noise at higher wavenumbers. The onset of instabilities is visible as the emergence of additional unstable modes that are not present at low strain values. For example, in Fig.~\ref{fig:instabilities}(b) additional unstable modes appear as a bright green color about $0.1 \lesssim k \lesssim 0.25$ mm$^{-1}$ for $\gamma \geq 3000-4000$. Additionally, in Fig.~\ref{fig:instabilities}(d) unstable modes appear as stippling for $\gamma \geq 1000$.} The onset of such instabilities are well beyond the time scales where transient flow reversals are observed. This applies to all fluids reported in this paper. The transient flow reversal occurs just beyond the stress overshoot and around $\gamma \approx 5$. Therefore, as discussed in our earlier work, it is unlikely that the onset of flow instabilities is related to the occurrence of transient flow reversals.
%


\subsection{\prevadded{Width of the shear stress plateau}}
\label{ssn:plateau-width}
\begin{figure}[h]
    \centering
    \includegraphics[width=0.5\textwidth]{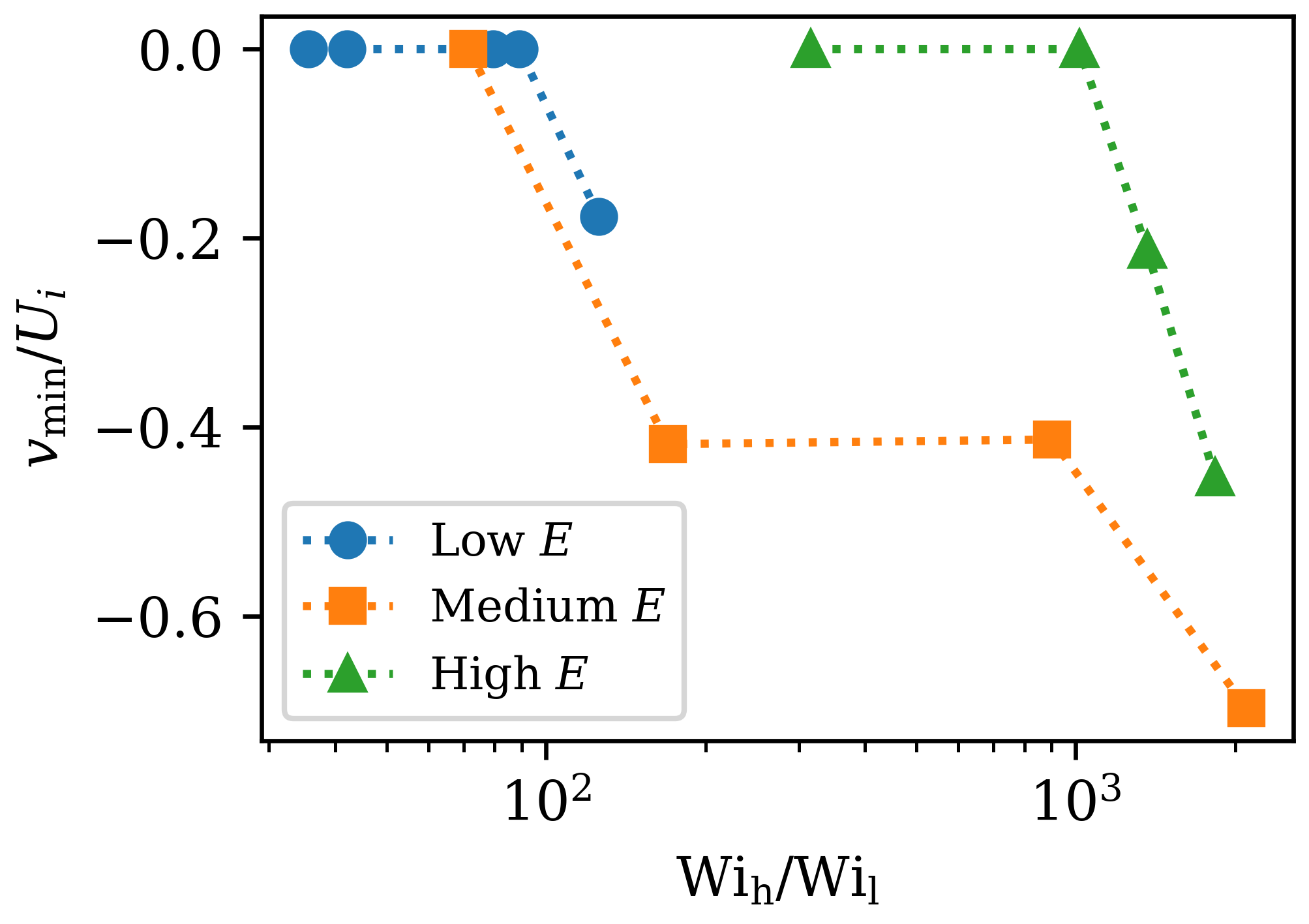}
    \caption{\prevadded{Dimensionless fluid velocities at the time and location of the fastest reverse flow for Wi~=~20 as a function of the stress plateau width. Here, Wi$_h$ and Wi$_l$ denote respectively the end and the onset of the stress plateau. }}
   \label{lowE_negativevelocity}
\end{figure}

\begin{figure}
    \centering
    \includegraphics[width=0.95\textwidth]{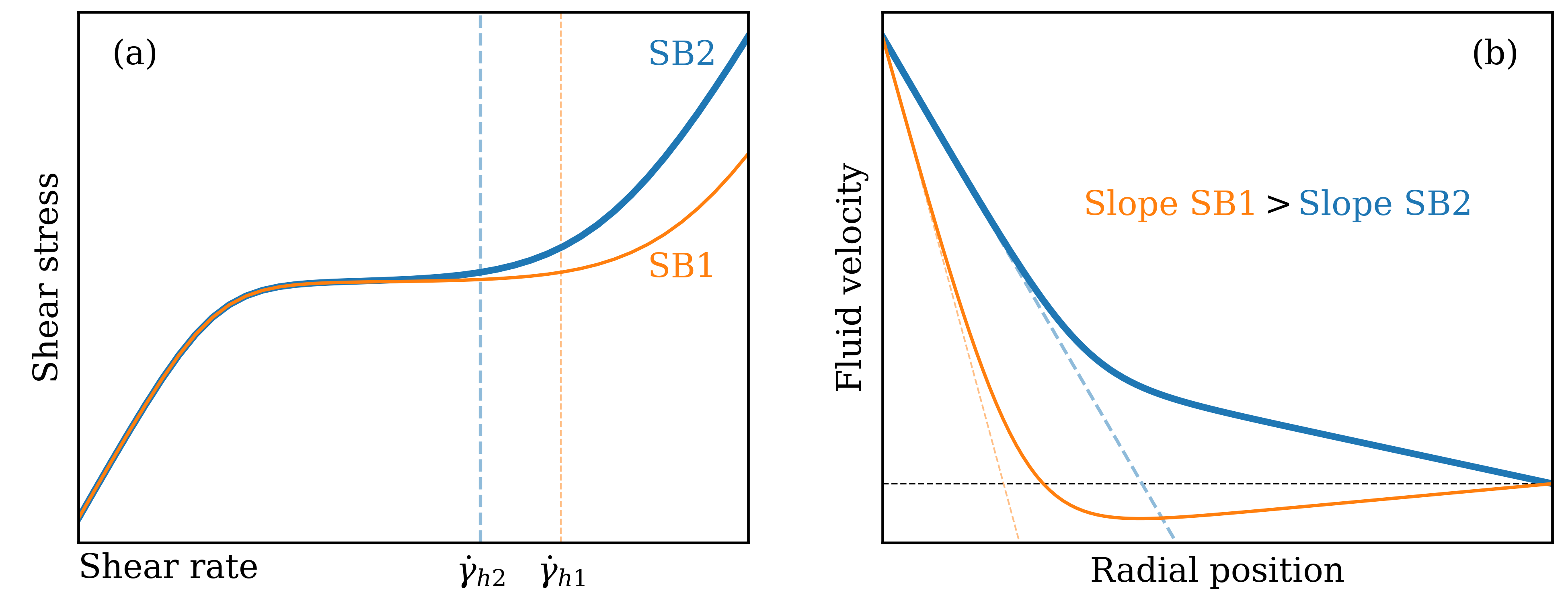}
    \caption{\prevadded{A schematic illustrating the proposed mechanism for the effect of shear stress plateau width on extent of transient flow reversal. (a) Sketch of flow curves for two WLM solutions with different plateau widths. (b) Sketch of corresponding velocity profiles when strongest flow reversal is observed.}}
    \label{fig:plateau-width-schematic}
\end{figure}

\prevadded{\prevadded{These experiments have revealed a non-monotonic trend in the occurrence or non-occurrence of the transient flow reversal with respect to the micellar entanglement \prevadded{number}.} As the micelles become more entangled and the entanglement number passes through a critical transition, the transient flow begins to exhibit reversals, but beyond a second critical threshold of the micellar entanglement \prevadded{number}, the flow reversal is not observed. Interestingly, there exists a correlation between the micellar entanglement \prevadded{number}, the width of the stress plateau, and the occurrence or not of the transient flow reversal. Fig.~\ref{lowE_negativevelocity} shows the greatest extent of the transient negative velocity as a function of the width of the stress plateau in experiments. Within each set of fluid preparations with fixed $E$, the preparations that exhibit transient flow reversal always have wider shear stress plateaus than those cases that do not show transient flow reversal. Additionally, for the sets with medium and high $E$, transient flow reversals become stronger with increasing plateau width.} \prevadded{The connection between the transient flow response and the width of the stress plateau may be rationalized as follows. Previous studies on shear banding WLMs \cite{Lerouge2008-interface-instabilities,Mohammadigoushki2016-interface-instabilities} have shown that upon imposition of the startup shear flow within the shear banding regime, the flow inside the gap of the TC cell undergoes a series of transitions. First, a linear velocity profile develops across the gap. Then, a high shear band forms near the rotating inner cylinder. A kink in the velocity profile forms at the juncture of the high and low shear rate bands. In time the velocity at the kink may overshoot to negative velocity values (see for example, Fig.~\ref{fig-hi-e-summary}(b,c)) before settling to the steady state value. This overshoot behavior is known as elastic recoil. At a fixed zero shear viscosity, systems with a wider stress plateau \add{exhibit a stronger decrease in the viscosity of the high shear band from $\eta_0$ in the quasi-steady flow. If the same trend applies to transient flow following the shear stress overshoot, such systems with a wider stress plateau} may have a higher \add{transient} shear rate (slope in the velocity) in the high shear band close to the inner cylinder (see a schematic shown in \prevadded{Fig.~\ref{fig:plateau-width-schematic}}). As the slope of the velocity near the inner cylinder increases \add{and the transient viscosity in this region decreases, the transfer of momentum from the inner cylinder to fluid beyond this region is interrupted to a greater extent. Consequently,} the likelihood of a strong elastic recoil, and thus the velocity reaching negative values during transient flow evolution, may also increase.} \add{Although the above argument is consistent with the experimental observations of this study, we note that it is based on the assumption that the slope of the high shear band during TFR is associated with the end of the stress plateau measured in quasi-steady experiments. This hypothesis is difficult to confirm experimentally and require higher spatial and temporal resolutions than accessible with our velocimetry technique. Therefore, further theoretical analysis of this hypothesis is warranted. }

As mentioned in the introduction, the DRP model predicted the correlation between the transient flow reversal and the entanglement \prevadded{number} \cite{Adams2011-DRP}.  The simulations of the transient flow of the DRP model also suggest that a stronger transient flow reversal is expected for wider stress plateaus. However, as the micellar entanglement \prevadded{number} $Z$ increases, the DRP model shows that the width of the plateau keeps widening monotonically which is different from the observation of the experimental data of this paper. \prevadded{One possible reason for this discrepancy is that the DRP model predicts stronger flow reversals for increasing values of the convective constraint release (CCR) parameter. To the best of our knowledge, this parameter cannot be quantified or controlled experimentally. Therefore, the observed non-monotonicity in the stress plateau width and transient flow reversals may be associated with uncontrolled changes in the CCR parameter. Preliminary simulations with the VCM model have also shown some similarities to experimental results in that there is non-monotonicity in the magnitude of the transient flow reversal with increasing $G'_{min}/G''_{min}$. The mechanism of this non-monotonicity is still unclear and will be addressed by simulations of the VCM model in the future.}\par
\prevadded{Despite the above attempt in connecting the equilibrium micellar entanglement number to the width of the stress plateau in WLMs, we note that the microstructure of the WLMs in strong non-linear flows is expected to be dynamic due to the feedback between flow and the microstructure. One should expect that due to potential occurrence of the flow-induced micellar breakage in strong flows, the dynamic micellar entanglement number under steady shear flow, and especially in the emerging or steady high shear band, is different from that of estimated from equilibrium micellar properties. Therefore, the connection between the equilibrium micellar entanglement number and the stress plateau is far from understood, which presents a challenge for future theoretical modeling and simulations of the shear banding WLMs. Another important aspect of these experiments is the general relevance of fluid elasticity to the transient flow reversal phenomenon. Fluid elasticity is defined based on the equilibrium micellar properties (zero shear viscosity and the relaxation time). As noted above, these equilibrium micellar properties change during highly non-linear transient flow experiments. As a result, the correlation between fluid elasticity and the transient flow reversal requires further analysis, which will be reported in our future studies.}

\section{Conclusions}
\label{scn-conclusions}

In summary, we have studied the impact of the micellar entanglement \prevadded{number} on the kinetics of shear banding flow formation in a range of shear banding wormlike micellar solutions. Our experiments show two critical transitions. First, as entanglement \prevadded{number} increases for a fluid set with a fixed elasticity number, we observe the emergence of a transient flow reversal during the shear stress decay. This is consistent with our prior observations \cite{Rassolov2020} and with DRP model predictions reported in the prior literature~\cite{Adams2011-DRP}. Surprisingly, beyond a second critical transition, the flow ceases to exhibit this transient flow reversal. Therefore, the observed extent of transient flow reversal in shear banding flow formation depends on the entanglement \prevadded{number} in a nontrivial manner. To better understand the connection between the fluid entanglement \prevadded{number} and the transient flow evolution, we compared extent of the flow reversal in experiments with the width of the shear stress plateau and identified a correlation: the extent of the transient flow reversal increases as the width of the stress plateau increases. However, the nature of connection between the micellar entanglement \prevadded{number} and the width of the stress plateau remains to be understood and will be the focus of our future research.

\section{Supplementary Material}
See supplementary materials for additional figures on flow profiles and spatio-temporal plots.

\section{Acknowledgments}
We are grateful to Joseph Schlenoff (FSU Chemistry and Biochemistry) and Daniel Hallinan (FAMU-FSU College of Engineering) who have given us access to their labs for silanization of the Taylor-Couette cell. Diffusing wave spectroscopy experiments were performed using the LS RheoLab II instrument available in the Ramakrishnan lab (FAMU-FSU College of Engineering) that is supported by NSF CREST 1735968. This work is funded by NSF CBET CAREER 1942150. We are grateful to Peter Olmsted, Gareth McKinley, Geoffrey Reynolds, Lin Zhou and Pam Cook for many helpful discussions.

\bibliography{refs}

\end{document}